\documentclass{ws-p8-50x6-00}

\begin{document}

\title{EUV and soft X--rays from clusters of galaxies -- the `cluster
soft--excess' phenomenon}

\author{Massimiliano Bonamente}
\address{Osservatorio Astrofisico di Catania, Catania 95125, Italy}

\author{Richard Lieu}
\address{University of Alabama, Huntsville, AL 35899, U.S.A.}

\maketitle

\abstracts{Clusters of galaxies are strong emitters of EUV and soft 
X-rays substantially in excess of the contribution from 
the hot intra--cluster medium at X--ray temperatures.
After the initial discovery of EUV excess emission in the Virgo cluster,
the phenomenon was reported also for the Coma, A1795 and A2199 clusters.
Recent reobservations confirmed the finding, and further searches 
on a large sample of nearby galaxy clusters are clearly
revealing the cosmological impact of the discovery.}

\section{Introduction}
The launch of the first X--ray satellites 
brought the astounding discovery that the space between galaxies in clusters
is filled with a very hot (T $\sim 10^{7-8}$ K) and tenuous gas
which emits X-ray radiation by bremsstrahlung processes. 
It is now clear that a hot intra--cluster medium (ICM) is a 
general feature of clusters of galaxies, which carries a susbstantial fraction
of a cluster's total mass.

As free-free radiation in a hot plasma generates 
broad band emission, some extreme--ultraviolet (EUV) and soft X--ray
radiation from clusters of galaxies is expected as the low--energy
tail of the hot ICM radiation.
In recent years, the  {\it Extreme Ultraviolet Explorer} (Bowyer
and Malina 1991),
an explorer class mission of NASA, opened a unique window to the 
EUV sky in the 65-200 eV passband~\footnote{With the {\it Deep Survey} Lexan/Boron
filter}, revealing that many clusters of galaxies 
emits EUV radiation substantially in {\it excess} of the expected contribution
from the hot plasma. The Virgo (Lieu et al. 1996a) and 
Coma clusters (Lieu et al. 1996b), two among the nearest clusters 
of galaxies that lie along lines of sight of low Galactic
hydrogen column densities ($N_H$),
were the first to show this {\it cluster soft--excess} (CSE)
 phenomenon, later followed by 
A1795 (Mittaz et al. 1998) and A2199 (Lieu, Bonamente and Mittaz 1999).

In this paper we highlight some of the main features of the CSE
by reporting the analysis and interpretation of recent EUVE re--observations
and of ROSAT PSPC data. Although the cause of the phenomenon
is still actively under investigation, we indicate  the possible origin of 
the emission, and comment on the cosmological impact of the CSE.

\section{EUVE observations with {\it in situ} background measurements}
The EUVE detectors are
known to have a time--variable background which originates
from several sources (Lieu at al. 1993).
It was recently shown (Lieu at al. 1999a) that the most appropriate way   
of handling the EUVE DS background is via an {\it in situ} measurement, 
i.e. a background measurement from a portion of sky devoid of
celestial sources which is placed at small off-set angles from the 
source under investigation, the measurement being carried out
time-contiguously to the source pointing.

\begin{figure}[h]
\psfig{file=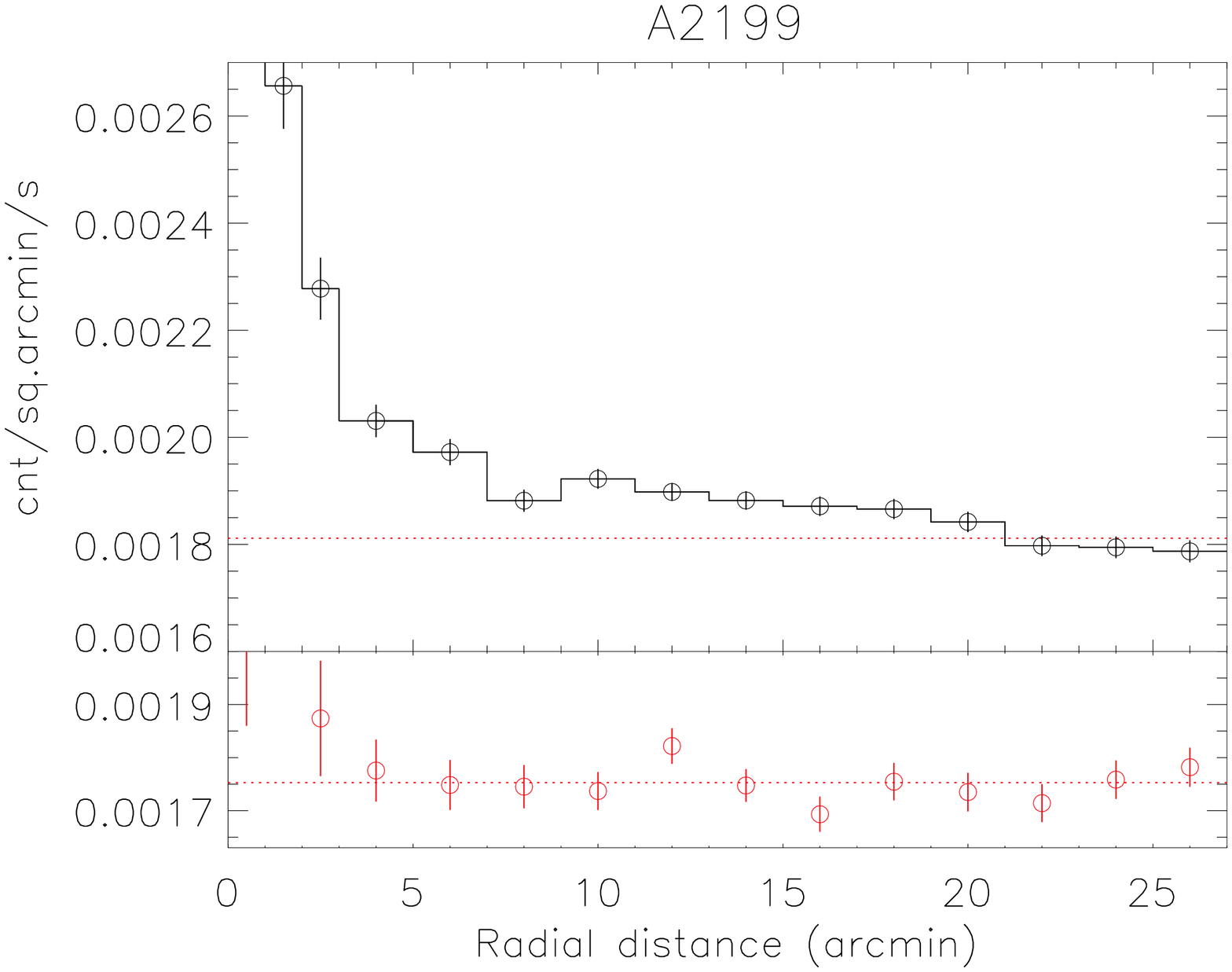,angle=90,height=4cm}
\psfig{file=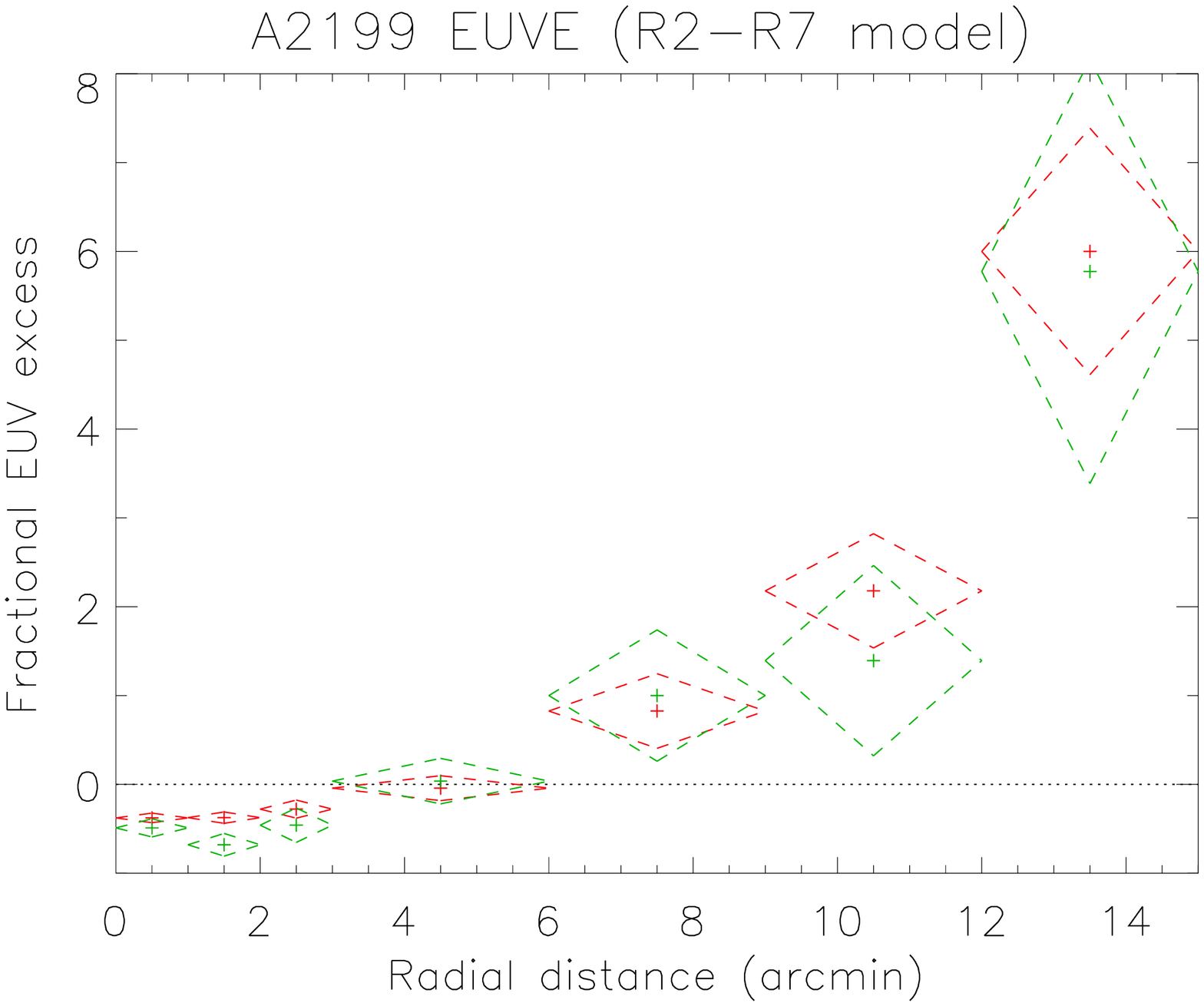,angle=90,height=4cm}
\caption{Left (a): radial profile of 
DS Lexan/Boron surface brightness of A2199 (top) and
the {\it in situ} background (bottom). Right (b): radial profile
of DS fractional excess, defined as $\eta = (M-E)/E$, where $M$ is measured
surface brightness and
$E$ is the expected value of the same, 
based on analysis of PSPC R2-R7 band (see 
footnote $b$ for details). 
The two sets of diamonds correspond respectively to the cases where
the best-fit background straight line (red) is subtracted from the data, 
and when a point-to-point background subtraction (green) is performed; the two
methods yield consistent results.
\label{fig:a2199}}
\end{figure}
 
Abell 2199 was the first cluster re-observed by EUVE in the forementioned
manner (Lieu 
et al. 1999a) , with resulting EUV emission detected out to
a radial distance of 20 arcmin (Fig. 1a), confirming the early findings.
When compared to the emission expected from the hot ICM~\footnote{
This is a relatively straightforward procedure.
For more detailed 
information, we refer the reader to Lieu et al. 1996a; Lieu, Bonamente
and Mittaz 1999; Mittaz et al. 1998; 
Bonamente, Lieu and Mittaz 2000a and references therein.  Essentially
we model the PSPC spectra (R2-R7 bands, or $\sim$ 0.2-2.0 keV by photon energy)
with a thin plasma emission code 
appropriate to the hot ICM, and calculate the expected fluxes from this
component in soft X-rays and EUV.} the CSE was found to be relatively more
prominent at large cluster radii (Fig. 1b): the surface brightness of the
CSE peters out with radius less rapidly than that of the hot ICM emission.
(the latter follows the customary
$\beta$ profile of intensities, see e.g. Cavaliere and Fusco--Femiano
1978).
A similar trend, to which we shall refer hereafter as 
the {\it soft--excess radial trend} (SERT), is also seen in Virgo and A1795,
as follows.

\begin{figure}[h]
\psfig{file=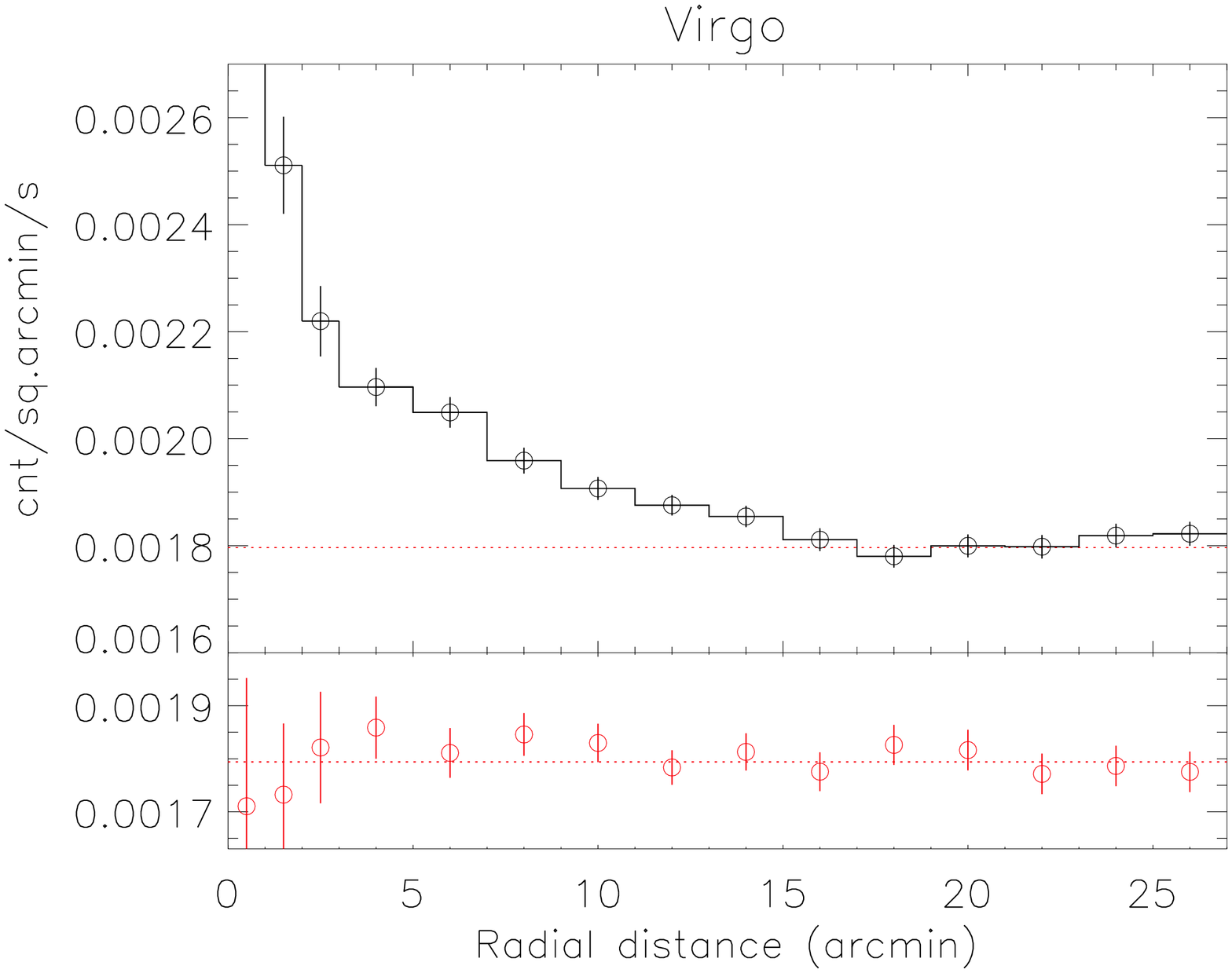,angle=90,height=4cm}
\psfig{file=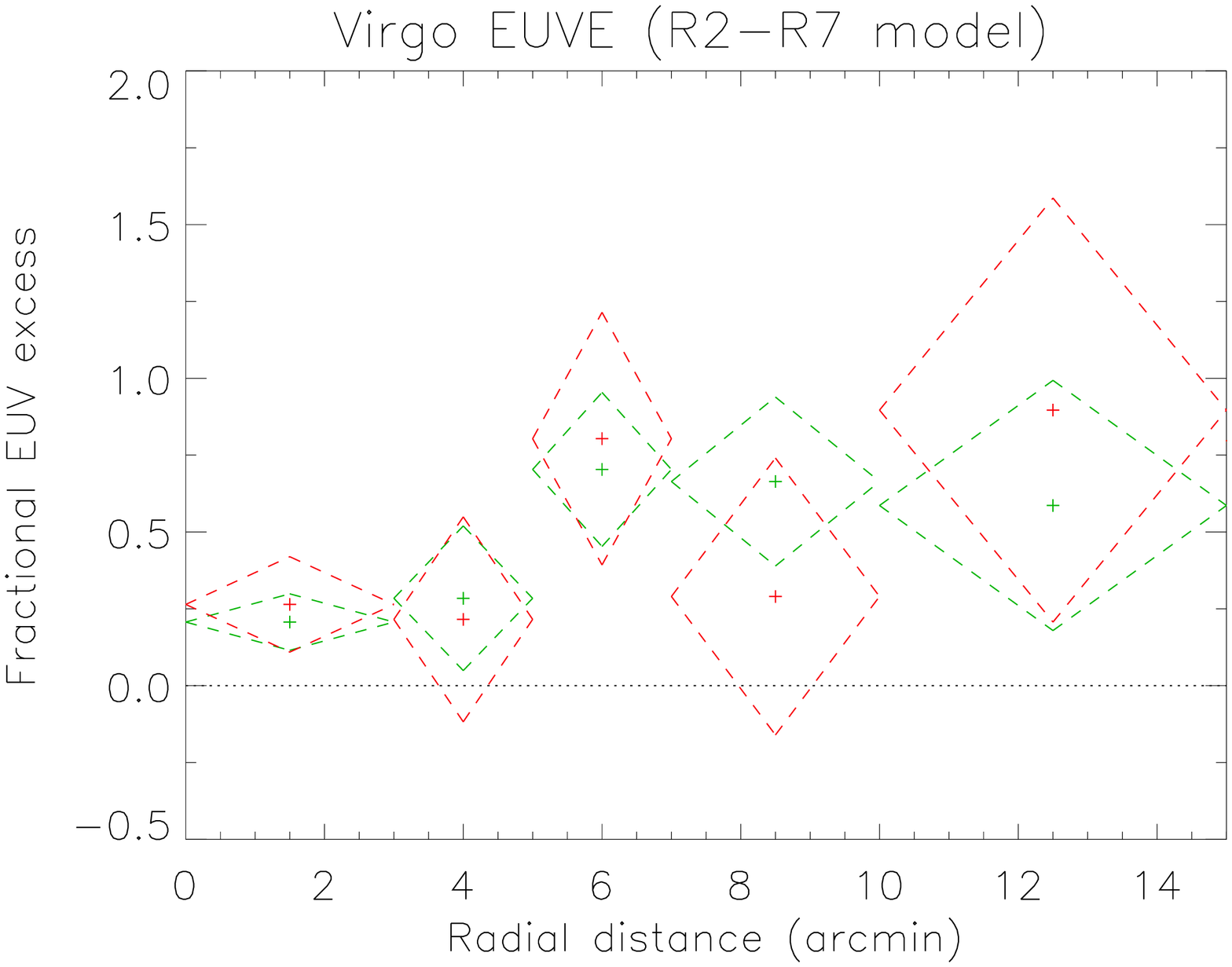,angle=90,height=4cm}
\caption{EUV surface brightness (left) and fractional excess (right) of Virgo;
see Fig. 1 for details.
\label{fig:virgo}}
\end{figure}

\begin{figure}[h]
\psfig{file=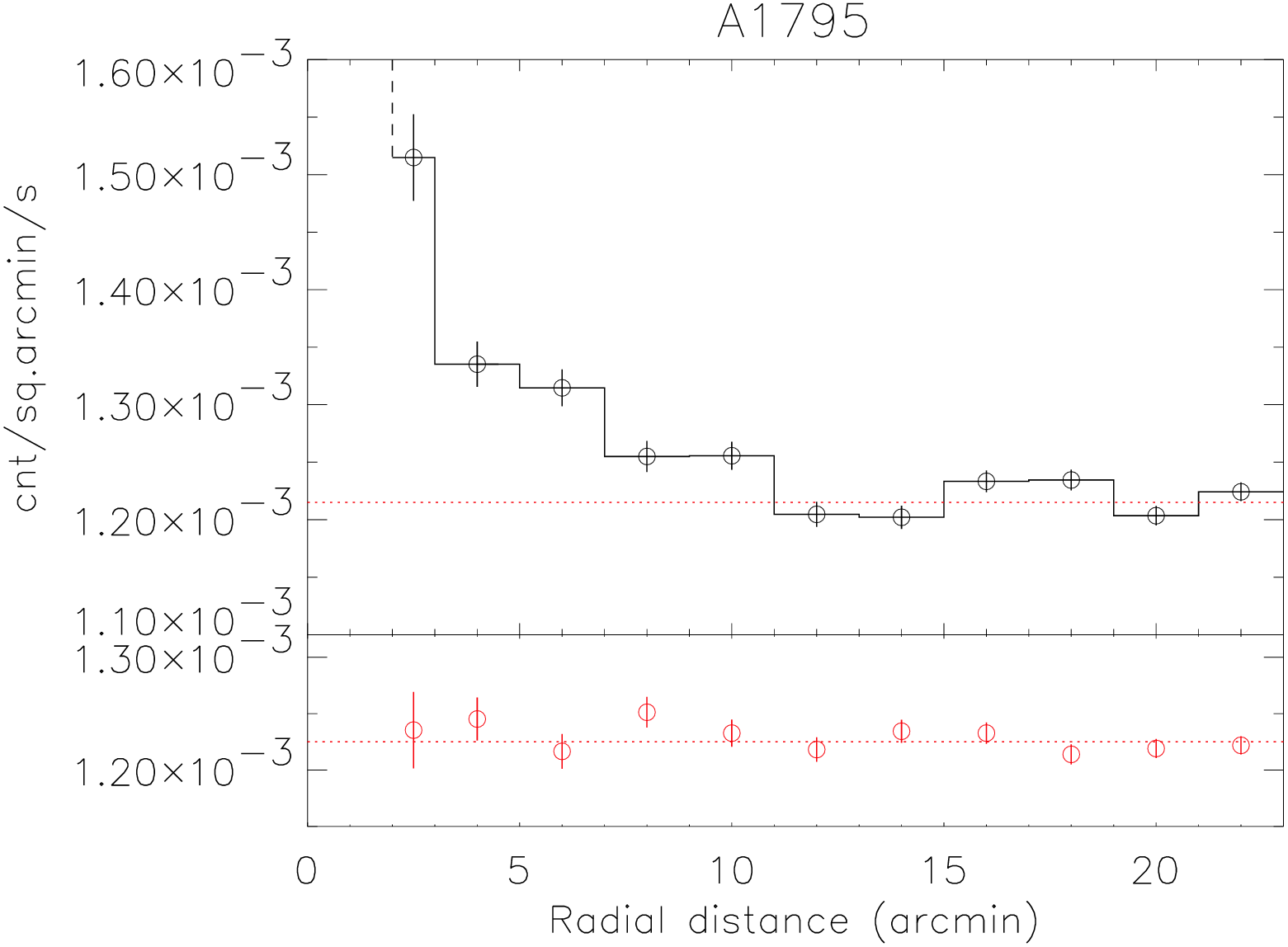,angle=90,height=4cm}
\psfig{file=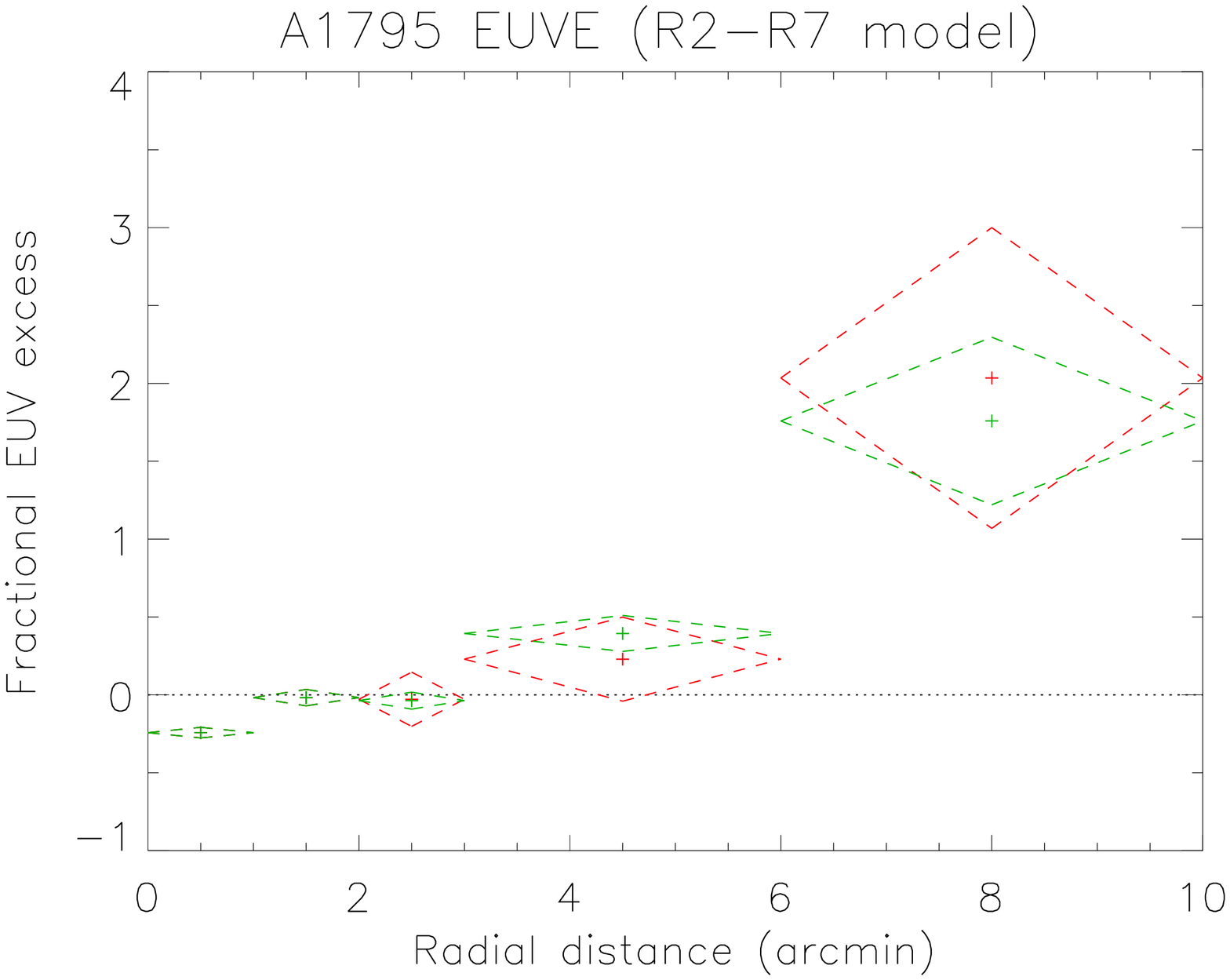,angle=90,height=4cm}
\caption{EUV surface brightness (left) and fractional excess (right) of A1795;
see Fig. 1 for details.
\label{fig:a1795}}
\end{figure}

EUVE reobservations were also performed on the A1795
and Virgo clusters with pointings (Bonamente, Lieu
and Mittaz 2000a). Following
the analysis method (of Lieu at al. 1999a) previously applied to A2199,
the Virgo cluster showed a halo of diffuse EUV emission out to a
radial distance of about 15 arcmin (Fig. 2a), and A1795
was detected out to a radius of about 10 arcmin (Fig. 3a), confirming
the early results.
The radial profiles of the excess emission for these two
clusters, see Fig. 2b and Fig. 3b, show a clear signature of the SERT 
effect.  

\section{Soft excess emission in the PSPC passband}
The ROSAT PSPC instrument has significant effective areas
at soft X--ray energies ($\sim 0.2-0.4$ keV, 
the so called R2 or C-band,
pulse--invariant channels 20-41; Snowden et al. 1994).
It is therefore possible to 
explore the presence of excess soft X--ray emission from clusters
of galaxies in a band that is neighboring to that of the EUVE {\it Deep Survey} 
instrument. This was already clear since the early investigations of
the CSE, as Lieu et al. (1996a, 1996b) reported detections of 
1/4 keV band excess emission from Coma and Virgo.
Later, also two pointed PSPC observations of  A1795 revealed
similar excesses at soft X--ray energies (Bonamente, Lieu and
Mittaz 2000a). In Fig. 4 we summarize the data.

\begin{figure}[h]
\psfig{file=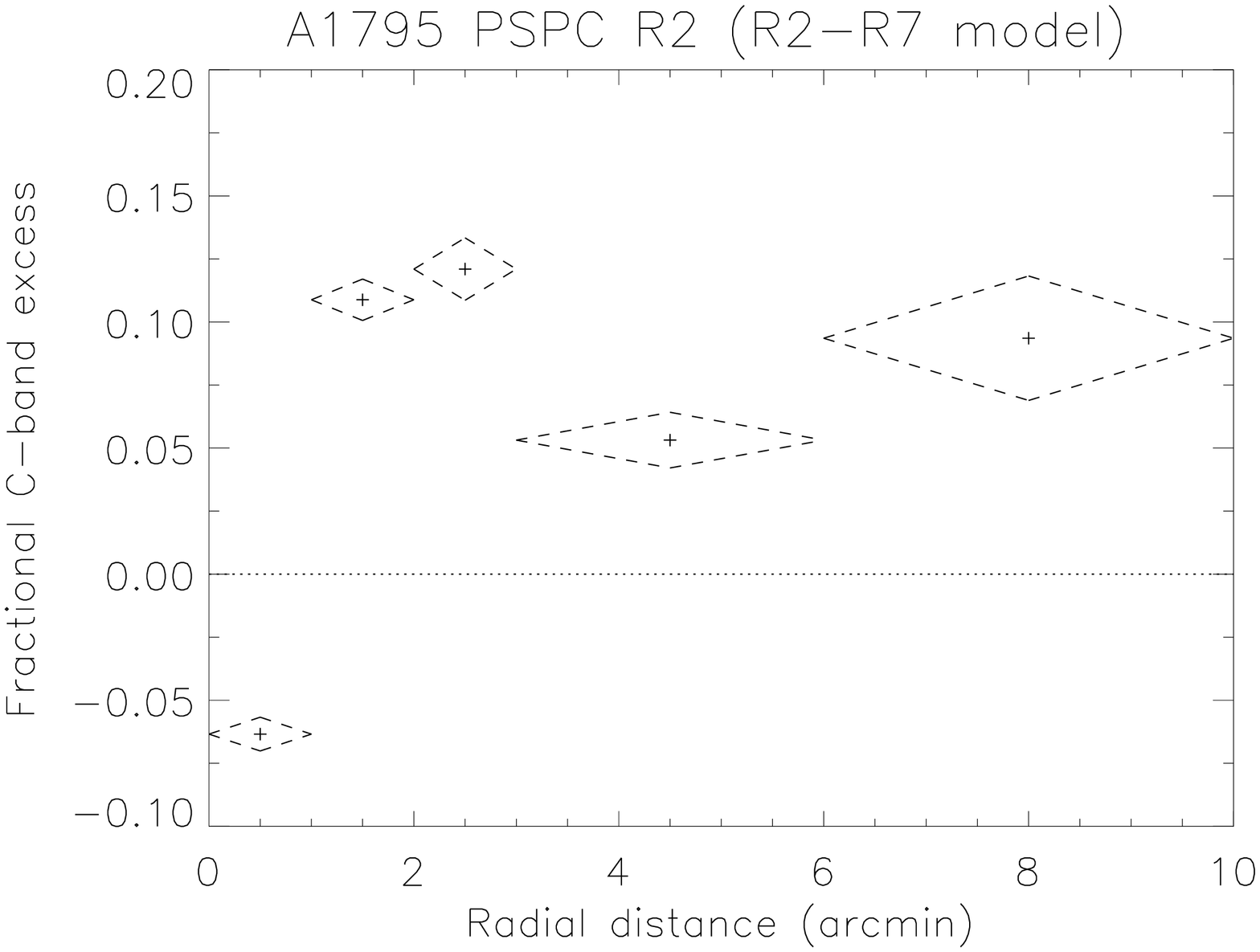,angle=90,height=4cm}
\psfig{file=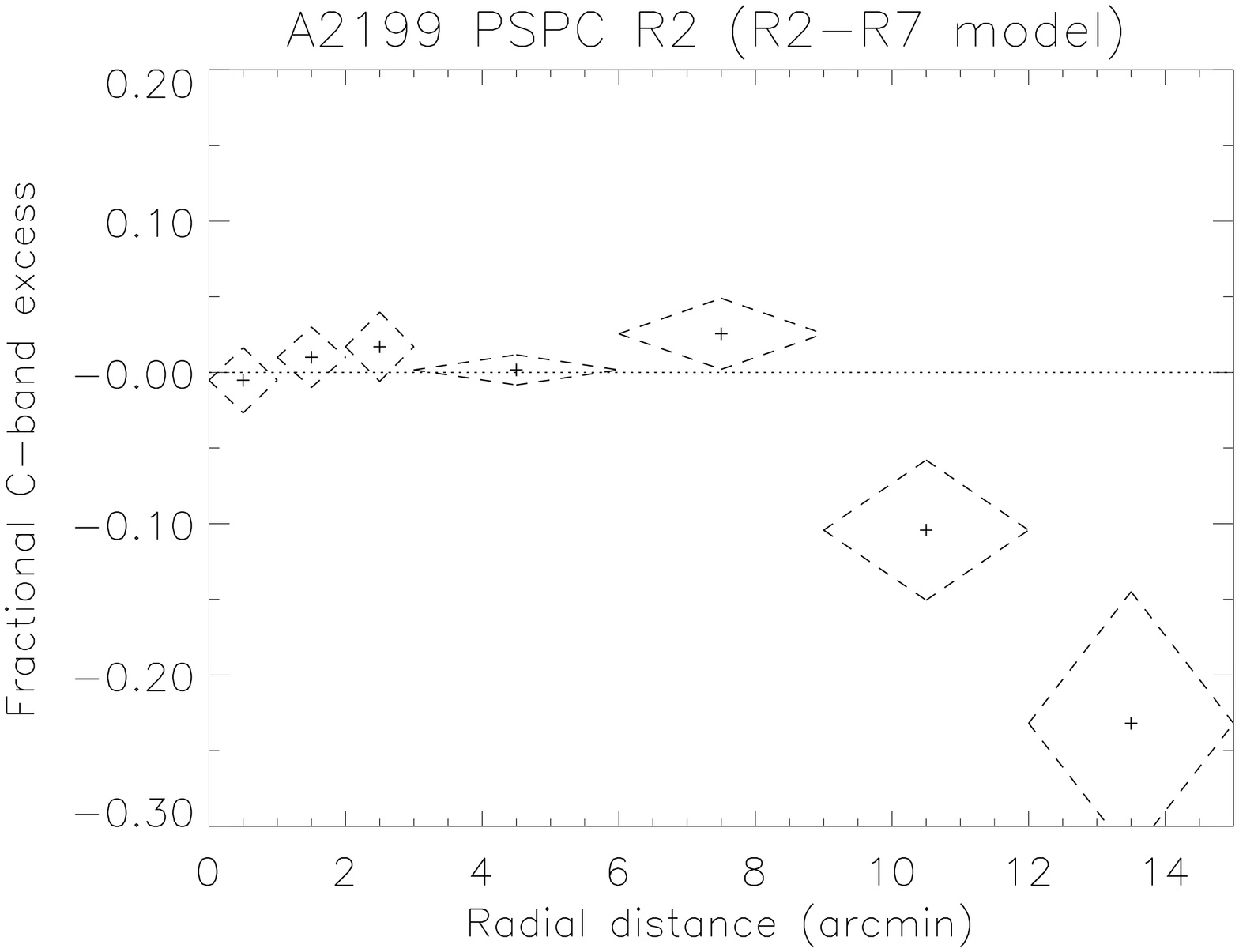,angle=90,height=4cm}
\psfig{file=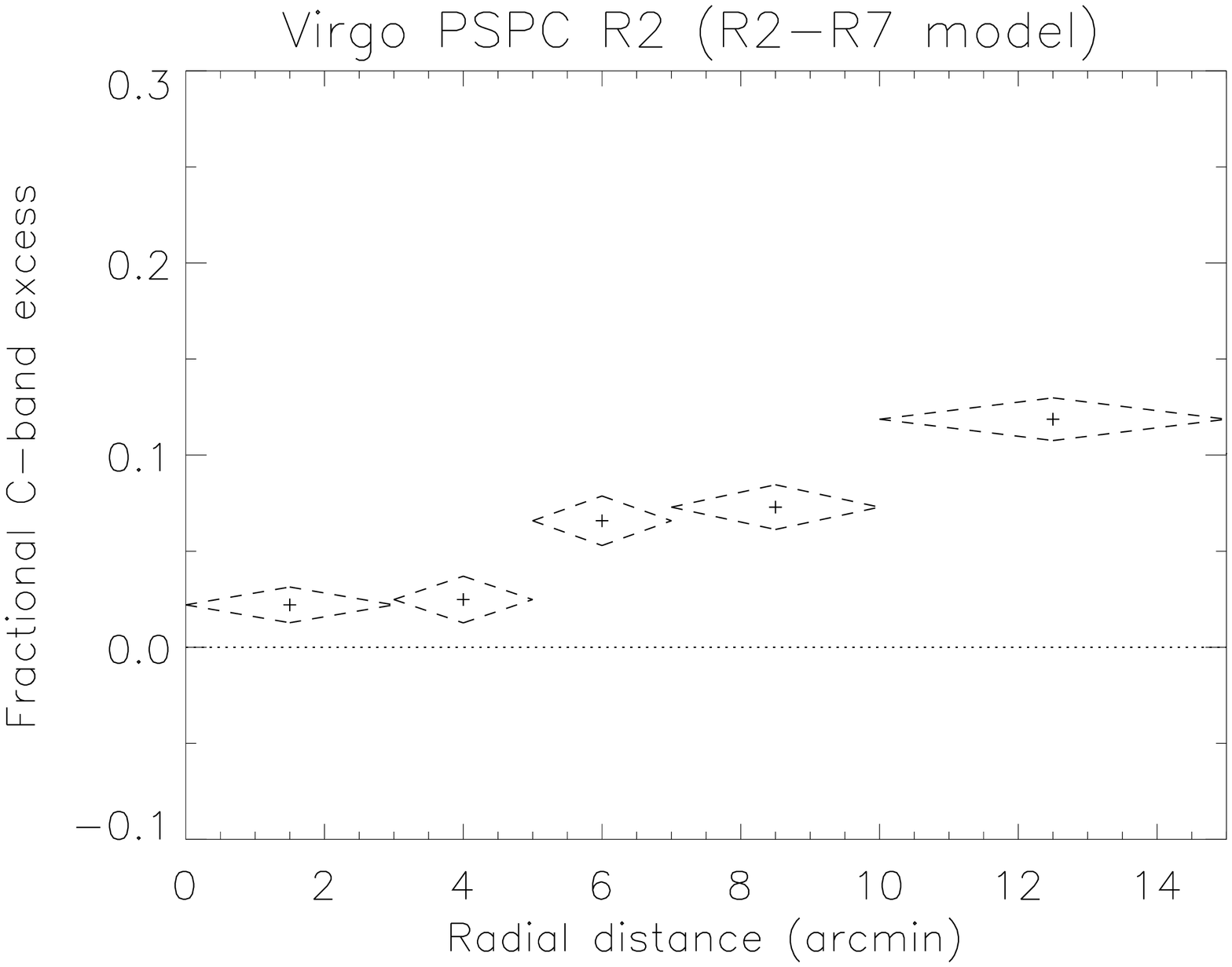,angle=90,height=4cm}
\psfig{file=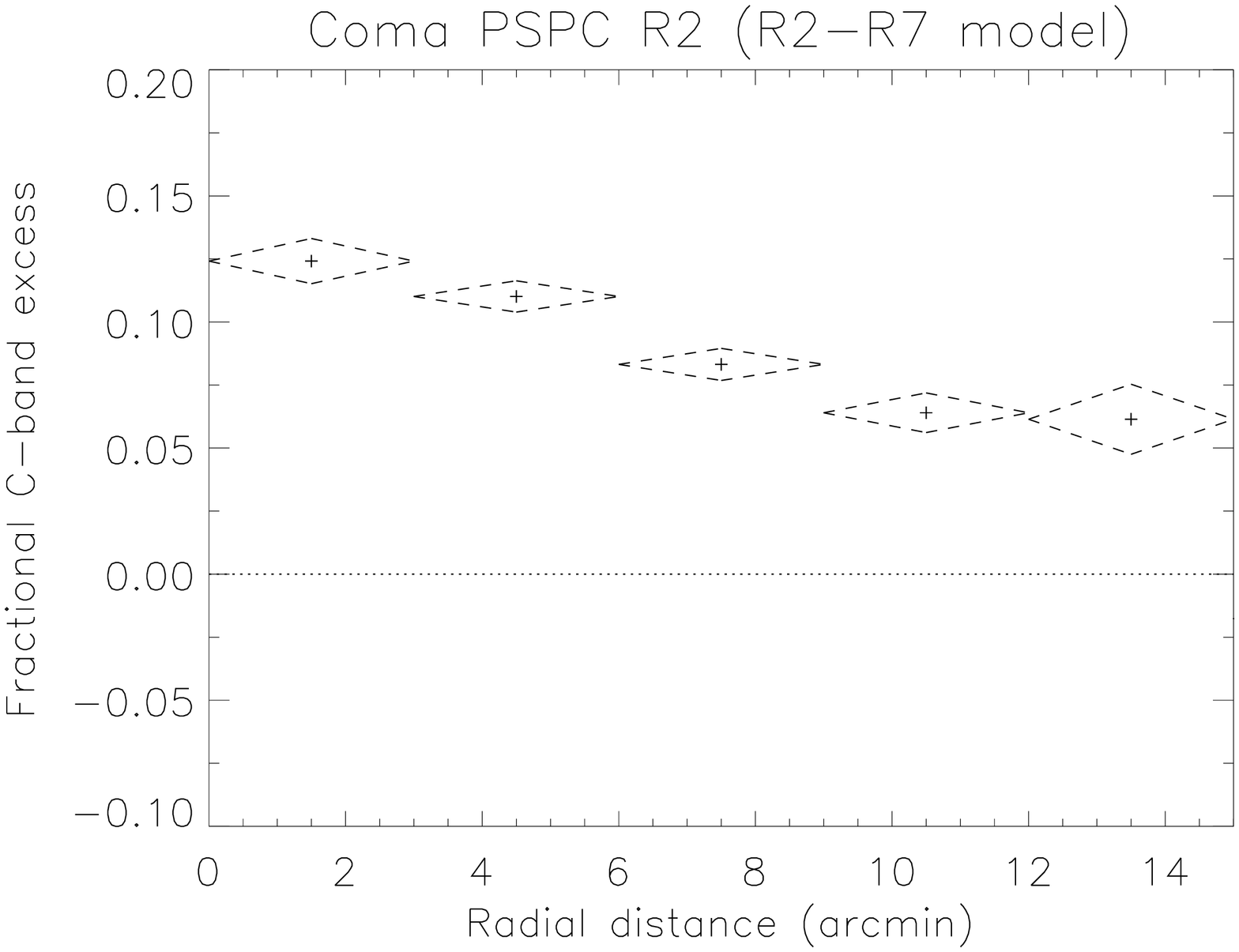,angle=90,height=4cm}
\caption{Fractional CSE plots in PSPC R2 band, $\sim$ 0.2-0.4 keV by
photon energy. The effect 
is clearly evident in all clusters, with the exception
of A2199, which apparently has a softer component 
detected only in the EUV (Fig. 1b).
\label{fig:cband}}
\end{figure}

The employment of PSPC carried the advantage of a large database of
clusters to be searched for soft--excess  emission.
We started our search for soft X--ray excess emission in the PSPC
passband with a sample of four nearby clusters member of the
Shapley concentration: A3571, A3558, A3560 and A3562, in 
the redshift range z $\sim$ 0.05-0.058. A3571 and A3558 are rich
and hot clusters (Markevitch et al. 1998), while A3560 and A3562 show
a lower X--ray luminosity.
The Shapley concentration lies along lines of sight of moderate
Galactic hydrogen ($N_H \sim$ 4-4.5 $\times 10^{20}$ cm$^{-2}$),
which absorbs about 95-99 \% of source soft X--ray photons.
This results in observations of moderate S/N, particularly for
the lower intensity clusters; four pointed PSPC
observations showed nonetheless clear evidence for
soft excess emission among {\it all} of these clusters, 
and particularly evident in A3571 and
A3558 (Fig. 5).  The 
detection of excess soft X--ray emission in the Shapley supercluster sample
indicates that CSE may be a common syndrome of
galaxy clusters.
A more detailed investigation of
these and other clusters
is in progress (Bonamente et al. 2001).

\begin{figure}[h]
\psfig{file=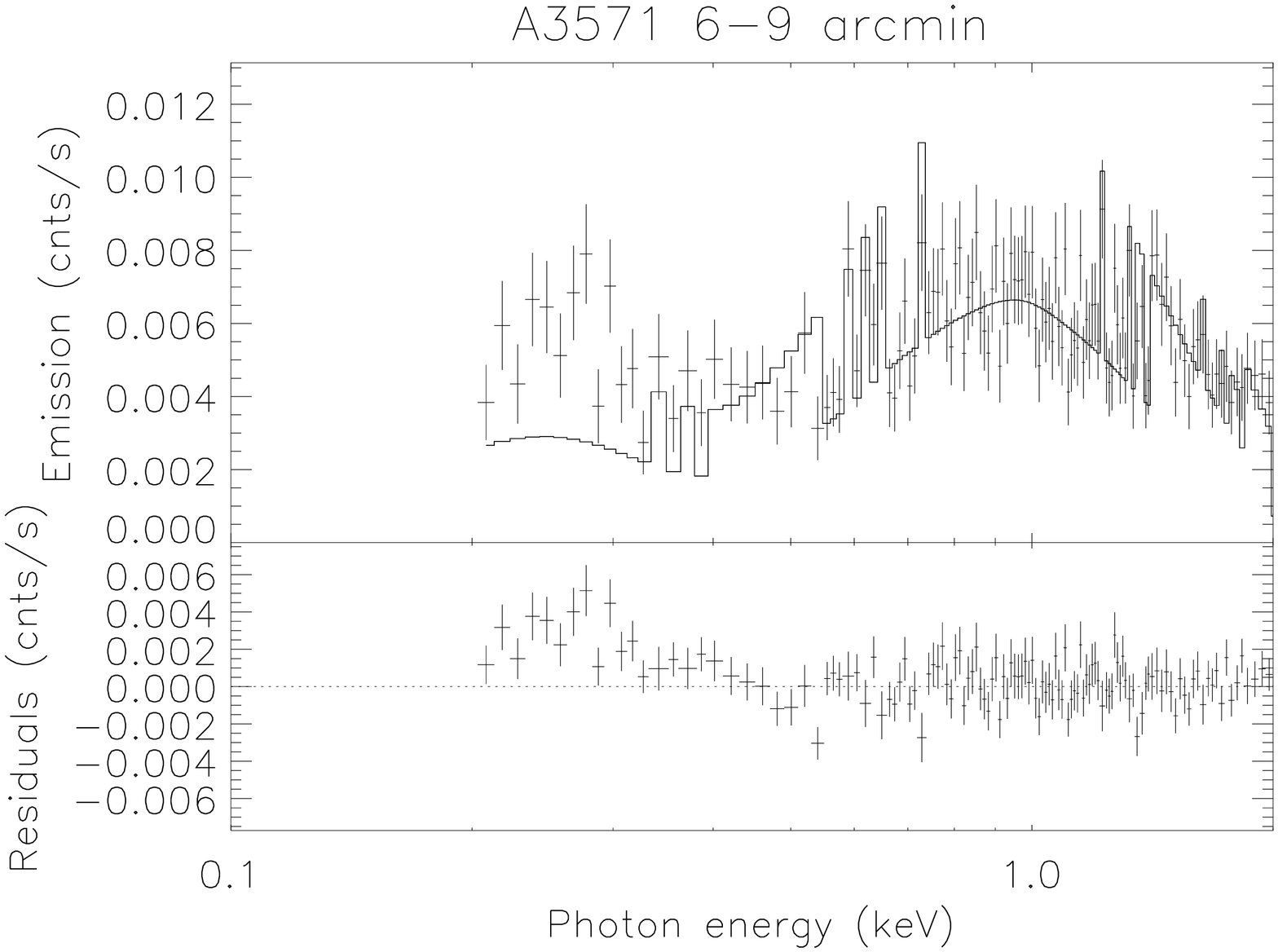,angle=90,height=4cm}
\psfig{file=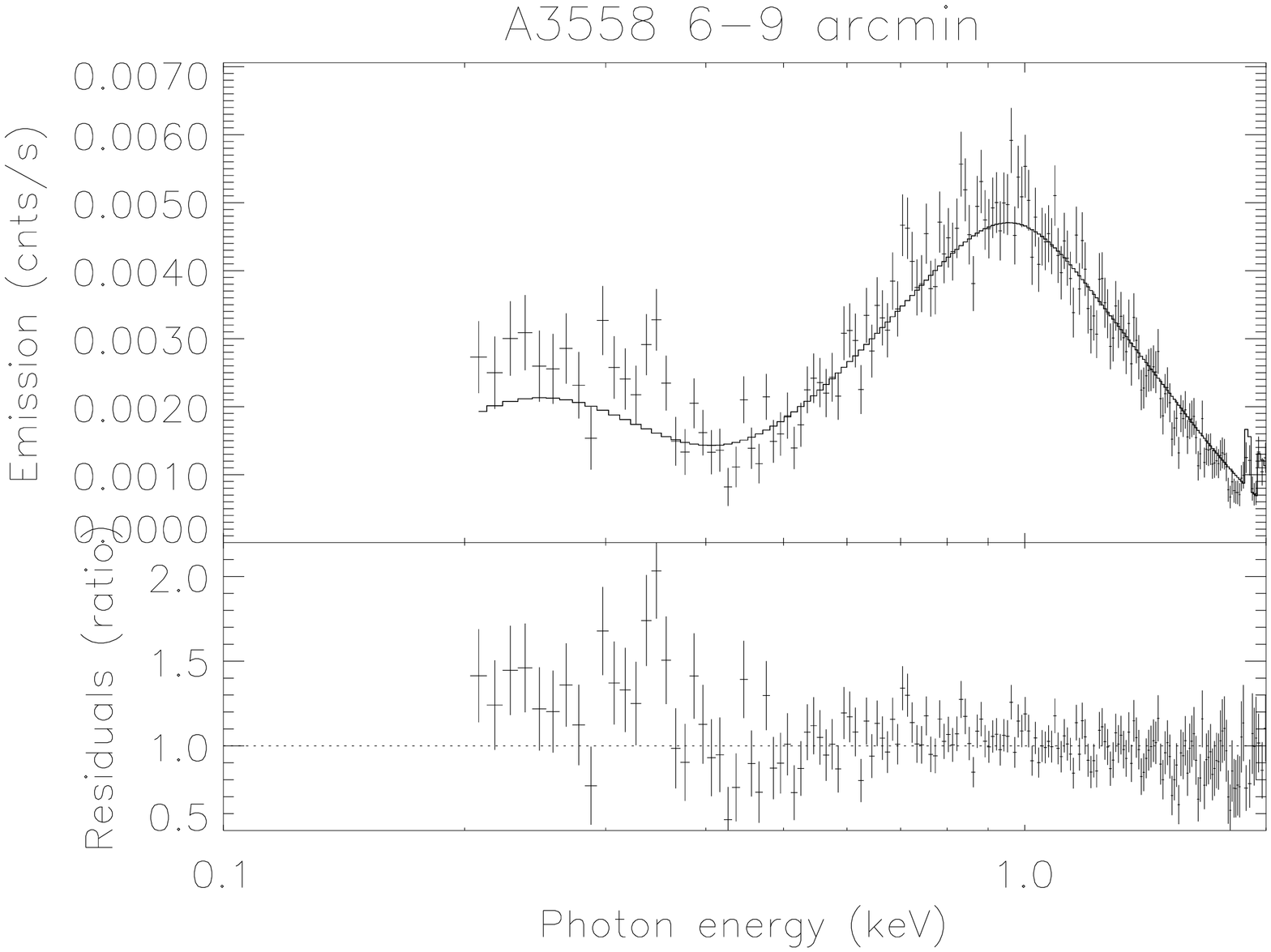,angle=90,height=4cm}
\caption{PSPC spectra of two regions of A3571 (left) and A3558 (right).
Excess above the best-fit thermal model is evident at the low-energy end
of the two spectra.
\label{fig:shapley}}
\end{figure}

\section{Interpretation of the excess fluxes}
Currently there are two main scenarii which contend for the interpretation
of the excess fluxes: the non--thermal model (Sarazin and Lieu 1998; Hwang
1997; Ensslin and Biermann 1998), which invokes an Inverse Compton effect
as cause for the emission, and the original thermal model (Lieu et al.
1996a) which invokes `warm' intra--cluster gas (at T$\sim 10^6$ K).
Although the issue may ultimately be resolved by 
spectroscopy, energetic requirements of the
non--thermal scenario appear quite demanding. For the case of the Coma
cluster (Lieu et al. 1999b), a population of
cosmic rays (CR) in approximate {\it energy
equipartition} with the hot ICM was enlisted to explain
the detected signals.  The requirements appear even more severe
in the case of A1795, and
the excess emission of Virgo meets no compelling interpretation
as an IC effect (Bonamente, Lieu and Mittaz 2000a).
On the other hand, the CSE may be interpreted as
radiation from a thin plasma at `warm' temperatures
(T$\sim$ 0.1 keV).
In more quantitative terms,
we provide in Tables 1--4  a comparison between the two interpretations
for Virgo and A1795, where we also entertained the notion
of `aging' of the relativistic electrons for the non--thermal model.

\begin{table}[h!]
\begin{center}
\caption{Modelling Virgo spectra with a two--temperature MEKAL (thermal
thin--plasma emissivity)  code (left),
MEKAL + power-law
model (center) and MEKAL + `aged' power-law  model (right). Errors
are 90 \% confidence.
 For the latter, the IC
magnetic field was fixed at 1$\mu$G,
the density of the ICM calculated from the best-fit $\beta$-model
 and
the electron differential number index $\gamma$ fixed at the
value of 2.5 (wich corresponds to a similar photon index $\alpha$ of 1.75).
The parameters of the hot phase were fixed at those best values obtained
by fitting PI channels 42 -- 201 with a single temperature model of
floating abundance.
$N_{warm}$ is here and after in
units of $10^{-14}$/4$\pi D^2 n^2 V$, where $D$ (cm) is the distance to the source,
$n$ (cm$^{-3}$)
is the gas density and $V$ (cm$^3$) is the volume of the emitting region.
For MEKAL+MEKAL and MEKAL+PO model fits, reduced $\chi^2$ is always between
1.1 and 1.3 for 180--181 d.o.f.
The non--thermal fits (MEKAL+PO and MEKAL+`aged' PO) are quite unsatisfactory.
The former returns a large value of $\alpha$, in disagreement with the
value of $\sim$1.75 (expected from a population of CR electrons with 
$\gamma \sim 2.5$), the latter 
shows large reduced $\chi^2$ values.
\label{tab:table1}}
\vspace{22pt}
\small
\begin{tabular}{cccccc}
\hline
 & \multicolumn{2}{c}{MEKAL+MEKAL} & \multicolumn{1}{c}{MEKAL+PO}
& \multicolumn{2}{c}{MEKAL+`aged' PO}\\
Region & $T_{warm}$ & $N_{warm}$
& $\alpha$ & $t_{age}$ & red. $\chi^2$ \\
 (arcmin) & (keV) & $\times 10^2$ & & (Gyrs) & (d.o.f.)  \\
\hline
0--3 & 0.066 $\pm^{0.038}_{0.014}$ & 0.36 $\pm^{0.24}_{0.16}$
& 3.3 $\pm^{2.4}_{0.6}$ & $\sim$ 0 & 1.6(181) \\
3--5 & 0.079 $\pm^{0.039}_{0.016}$ & 0.28 $\pm^{0.07}_{0.08}$
& 4 $\pm^{1.4}_{0.6}$ & $\sim$ 0 & 1.66(181) \\
5--7 & 0.079 $\pm^{0.022}_{0.015}$ & 0.4 $\pm^{0.2}_{0.12}$
& 4.4 $\pm^{1.2}_{0.7}$ & $\sim$ 1.1 & 2(180) \\
7--10& 0.09 $\pm^{0.01}_{0.012} $ & 0.59 $\pm^{0.25}_{0.13}$
& 4.2$\pm^{0.8}_{0.5}$ & 1.8$\pm^{0.015}_{0.02}$ & 1.11(180) \\
10--15&0.083 $\pm^{0.009}_{0.011}$ & 1.1 $\pm^{0.38}_{0.1}$
& 4.5$\pm^{0.65}_{0.3}$ & 2.1$\pm^{0.15}_{0.05}$ & 1.2(180) \\
\hline
\end{tabular}
\end{center}
\end{table}

\begin{table}[h!]
\begin{center}
\caption{
Modelling of A1795 spectra with a two-temperature MEKAL code (left) and
with a MEKAL + `aged' power-law code (right); errors are 90 \% confidence.
Parameters of the hot phase were
fixed at those best values obtained by fitting PI channels 42 -- 201 with
a single temperature model of abundance 0.31 solar.
For the non-thermal model, the value of the
IC magnetic field was fixed at 1$\mu$G; the density of the ICM, a parameter which
affects the aging of electrons, was calculated from the best-fit
$\beta$-model. Electron differential index $\gamma$ was fixed
at the value of 2.5 (see Table 1).
\label{tab:a1795_non}}
\vspace{22pt}
\small
\begin{tabular}{ccccccc}
\hline
 & \multicolumn{3}{c}{MEKAL+MEKAL} & \multicolumn{2}{c}{MEKAL+`aged' PO}  \\
Region  & $T_{ warm}$ & $N_{warm}$ &red. $\chi ^2$
  & $t_{age}$  & Red. $\chi^2$ \\
 (arcmin) & (keV)&$\times 10^2$ & (d.o.f.) &  (Gyrs)    & (d.o.f.)  \\
\hline
1--2 & 0.125$\pm^{0.034}_{0.037}$ & 0.045$\pm^{0.008}_{0.009}$ & 1.0(363)
& 1.7 $\pm^{0.17}_{0.11}$ &   1.0(363)\\
2--3 & 0.11$\pm^{0.04}_{0.02}$& 0.0285$\pm^{0.006}_{0.007}$ & 1.16(363)
& 2.25 $\pm^{0.15}_{0.55}$  & 1.15(363)\\
3--6 & 0.049$\pm^{0.017}_{0.015}$& 0.12$\pm^{0.37}_{0.06}$ & 0.97(363)
& 2.8 $\pm^{0.1}_{0.13}$ & 0.98(363) \\
6--10& 0.025$\pm^{0.013}_{0.012}$ & 110$\pm^{700}_{70}$ & 0.88(363)
& 3.7 $\pm^{0.16}_{0.14}$    &  0.88(363) \\
\hline
\end{tabular}
\end{center}
\end{table}

\begin{table}[h!]
\begin{center}
\caption{Luminosity and pressure estimates for A1795;
$L^{CSE}_{42}$ is the intrinsic luminosity of non--thermal best--fit
model between photon energies 65--250 eV ($\gamma_{min}$=279 and
$\gamma_{max}$=548) in units of 10$^{42}$ (erg s$^{-1}$) and
$P^{CSE}$ is the pressure of relativistic electrons
in the [$\gamma_{min}$,$\gamma_{max}$] Lorentz factor interval.
$P^{e}$ and $P^e(t=0)$ are
the total electron pressure at the present epoch
and at the acceleration epoch, respectively, assuming cosmological
parameters $H_0=75$ km s$^{-1}$ Mpc$^{-1}$ and $q_0=0$. If $H_0=50$ km s$^{-1}$ Mpc$^{-1}$
is adopted, all present pressure estimates for the
non-thermal component will increase $\sim$ two-fold.
Pressure for the ICM ($P^{gas}$) was calculated according
to the parameters of the best-fit $\beta$ model.
\label{tab:a1795_pr}}
\vspace{22pt}
\small
\begin{tabular}{cccccc}
\hline
Region & $L^{CSE}_{42}$ & $P^{CSE}$ & $P^{gas}/P^{CSE}$ &
$P^{gas}/P^e$ & $P^{gas}/P^e(t=0)$ \\
(arcmin) & & (erg cm$^{-3}$) & & &  \\
\hline
1--2 & 0.9 & 1.6 $10^{-13}$ & 88.3 & 35.3 & 1.9  \\
2--3 & 1.3 & 1.0 $10^{-13}$ & 128& 85.8 & 4.8  \\
3--6 & 4.5 & 1.0 $10^{-13}$ & 120 & 52.7 & 2.3   \\
6--10& 22.0& 6.8 $10^{-14}$ &  17.6& 2.5  & 0.22  \\
\hline
\end{tabular}
\end{center}
\end{table}

\begin{table}[h]
\begin{center}
\caption{Warm gas density estimates and comparison with the hot ICM for
the Virgo and A1795 cluster (see Tables 1 and 2). Warm gas densities ($n_{ warm}$)
are estimated for 100\% filling factors.}
\vspace{22pt}
\small
\begin{tabular}{cccccc}
\hline
\multicolumn{3}{c}{Virgo} & \multicolumn{3}{c}{A1795} \\
Region & $n_{ warm}$ & $
n_{hot}$ & Region & $n_{warm}$ & $n_{hot}$\\
(arcmin) & ($10^{-3}$ cm$^{-3}$) &($10^{-3}$ cm$^{-3}$)
& (arcmin) & ($10^{-3}$ cm$^{-3}$) & ($10^{-3}$ cm$^{-3}$) \\
\hline
0--3 & 5.2   $\pm^{3.4}_{2.4}$     & 50 & 0--1 & --- & 3.7 \\
3--5 & 2.4 $\pm^{0.6}_{0.65}$   &10  & 1--2 & 0.92$\pm^{0.16}_{0.18}$ & 3.3 \\
5--7 & 1.9 $\pm^{1}_{0.6}$   &7.5 & 2--3 & 0.45$\pm^{0.1}_{0.11}$  & 2.7 \\
7--10& 1.4 $\pm^{0.6}_{0.3}$   &5   & 3--6 & 0.3$\pm^{0.9}_{0.15}$  & 1.6 \\
10--15&1.0$\pm^{0.34}_{0.1}$ &3   & 6--10& 4.5$\pm^{26}_{2.6}$     & 0.93\\
\hline
\end{tabular}
\end{center}
\end{table}

If the CSE is indicative of `warm' baryons in clusters of galaxies,
what are then their mass implications?
In absence of a precise knowledge of the clumping of the warm gas,
a 100 \% filling factor results in masses comparable to those of the hot
phase (see Table 4 for the cases of Virgo and A1795), and
potentially even larger at large cluster radii, as 
indicated by the increasing trend (SERT) of the fractional
EUV excesses (Fig. 2 and 3). Similar mass implications also apply to the excess
emission of A2199 (see Lieu et al. 1999a).

The consensus on the baryonic content of our Universe, as reported by
many independent researchers also at this Conference, ranges between
$\Omega_b \simeq 0.017 - 0.045$, whereby the total matter
content is currently parametrised as $\Omega_M \simeq 0.25 - 0.45$.
Our results show that the cluster soft--excess phenomenon
may be indicative of a new baryonic component with mass density (relative to
the `critical' density)
$\Omega_{CSE}$ which could constitute a sizeable
fraction of $\Omega_b$, although probably not so massive as to bridge the gap
between the current estimates of $\Omega_b$ and $\Omega_M$. 
 
\section{Conclusions}
We showed that EUV and soft X--ray observations of clusters of galaxies 
reveal the presence of {\it excess} emission above the well known
radiation from the hot ICM. A possible interpretation
of this phenomenon invokes substantial quantities of {\it warm} gas that
may coexist with gases at higher and lower temperatures.
Recent investigations of PSPC images also provided evidence for
cold gas (T $\leq 10^5$ K) with small covering factors (Bonamente,
Lieu and Mittaz 2000b). As more clusters  are found to
exhibit the
phenomenon, its possible relevance to cosmology increases.
Could this be in the form of a new contribution to the baryonic content of our
Universe ($\Omega_{CSE}$) ?

\section*{Acknowledgments}
We acknowledge  NASA/ADP for financial support. MB acknowledges the support
of a post--doctoral fellowship at the Catania Astrophysical Observatory. 
RL is grateful to the IAP and Florence Durret for hospitality.

\end{document}